\documentclass[preprint]{revtex4-1} %twoside, 

\usepackage[dvipsnames]{xcolor}
\usepackage[utf8]{inputenc}
\usepackage[T1]{fontenc}
\usepackage{amssymb}
\usepackage[colorinlistoftodos, color=green!40, prependcaption]{todonotes}
\usepackage{amsthm}
\usepackage{mathtools}
\usepackage{physics}
\usepackage{diagbox}
\usepackage{makecell}
\usepackage[left=20mm,right=20mm,top=30mm, columnsep=15pt]{geometry}
\usepackage{adjustbox}
\usepackage{relsize}
\usepackage{placeins}
\usepackage{hyperref}
\usepackage{extarrows}
\usepackage{csquotes}
\usepackage[]{graphicx}
\usepackage{float}
\usepackage{qcircuit}
\usepackage{pgfplots}
\usepackage{amsmath}
\usepackage{braket}
\usepackage{multirow}
\usepackage{verbatim}
\usepackage{dsfont}
\usepackage{bm}
\usepackage[shortlabels]{enumitem}
\usepackage{rotating}
\numberwithin{equation}{section}

\definecolor{USI}{RGB}{254,226,124}
\definecolor{Wien}{RGB}{0,99,166}
\definecolor{QSIT}{RGB}{0,152,155}
\definecolor{qcol}{RGB}{82,41,125}
\definecolor{Bblue}{rgb}{0.19, 0.55, 0.91}

\newcommand{\proj}[1]{|#1\rangle\langle#1|}
\begin{document}
\title{Classical information and collapse in Wigner's friend setups}

\author{Veronika Baumann}
    \email{veronika.baumann@oeaw.ac.at}% Your name
     \affiliation{Atominstitut, Technische Universität Wien, 1020 Vienna, Austria}
    \affiliation{Institute for Quantum Optics and Quantum Information (IQOQI), Boltzmanngasse 3, 1090 Vienna, Austria}
    
\date{\today}

\begin{abstract}
The famous Wigner's friend experiment considers an observer -- the friend-- and a superobserver --Wigner-- who treats the friend as a quantum system and her interaction with other quantum systems as unitary dynamics. This is at odds with the friend describing this interaction via collapse dynamics, if she interacts with the quantum system in a way that she would consider a measurement. These different descriptions constitute the Wigner's friend paradox. Extended Wigner's friend experiments combine the original thought experiment with non-locality setups. This allows for deriving local friendliness inequalities, similar to Bell's theorem, which can be violated for certain extended Wigner's friend scenarios.
A Wigner's friend paradox and the violation of local friendliness inequalities require that no classical record exists, which reveals the result the friend observed during her measurement. Otherwise Wigner agrees with his friend's description and no local friendliness inequality can be violated. In this article, I introduce classical communication between Wigner and his friend and discuss its effects on the simple as well as extended Wigner's friend experiments. By controlling the properties of a (quasi) classical communication channel between Wigner and the friend one can regulate how much outcome information about the friend's measurement is revealed. This gives a smooth transition between the paradoxical description and the possibility of violating local friendliness inequalities, on the one hand, and the effectively collapsed case, on the other hand. %In the latter Wigner agrees with the friend's probability assignments and local friendliness inequalities cannot be violated.
\end{abstract}

\maketitle

%%%%%%%%%%%%%%%%%%%%%%%%%%%%%%%%%%%%%%%%%%%%%%%%%%%%%%

\section{Introduction}
\label{Introduction}

The Wigner's friend thought experiment was originally proposed in~\cite{wigner1963problem} to reason about the applicability of the two dynamics featured by quantum mechanics. On the one hand sufficiently isolated quantum systems evolve unitarily. On the other hand, quantum systems upon measurement undergo a seemingly instantaneous collapse to the eigenstate corresponding to the observed outcome. These two different dynamics and the lack of a clear prescription of when to use one or the other is one of the main aspects of the quantum measurement problem~\cite{bub2010two,busch1996quantum,maudlin1995three}. 

\begin{figure}[hbt!]
\begin{tikzpicture}[scale=0.75]

  \draw[thick,] (-4,-3) rectangle (5.5,3);
  \node[fill=RoyalPurple!30, draw=RoyalPurple, thick] (s) at (-3,0) {$S$};

   \draw[->,>=latex, thick,] (s) -- node [midway,below] {$\ket{\phi}_S$} (-0.5,0);
      
  \node[fill=Green!20, draw=Green, thick, rounded corners=3pt] (r1) at (3.75,1.75) {$\ket{0}_S\ket{\bm 0}_F\,$};
  \node[fill=Green!20, draw=Green, thick, rounded corners=3pt] (r2) at (3.75,-1.75) {$\ket{1}_S\ket{\bm 1}_F\,$};
  \draw[->,>=latex, thick, draw=gray] (1.25,0.75) to[bend left] (r1);
  \draw[->,>=latex, thick, draw=gray] (1.25,-0.75) to[bend right] (r2);
  
    \node[xscale=-1] (m) at (1,0) {\includegraphics[scale=0.035]{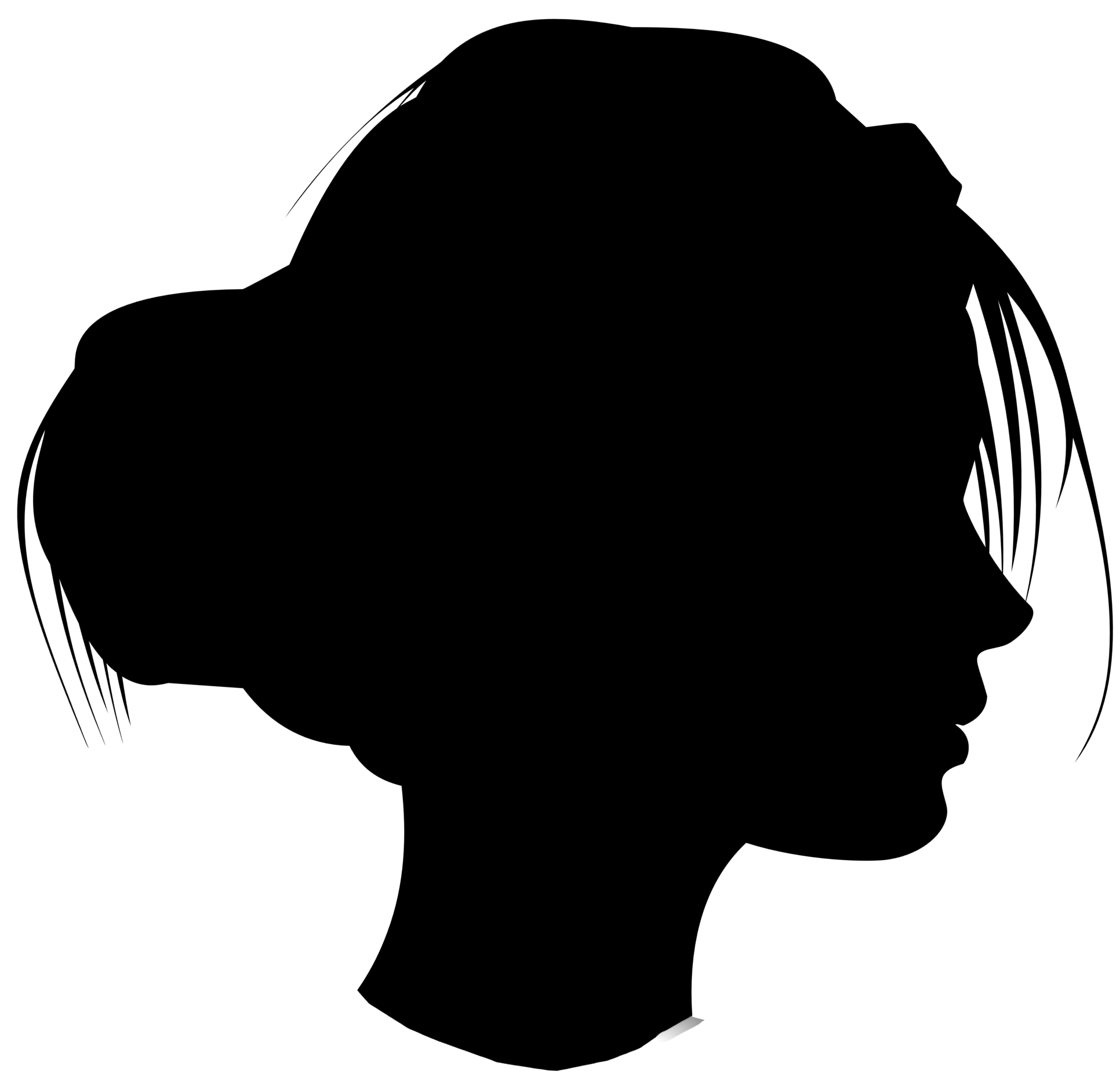}};
  \node[] (F) at (1,-2) {\textcolor{Green}F};

  \draw[->,>=latex, thick] (5.5,0) --node [midway,below]{$\ket{\Phi}_{SF}$} (7.75,0);

  \node[] (M) at (9.125,0) {\includegraphics[scale=0.075]{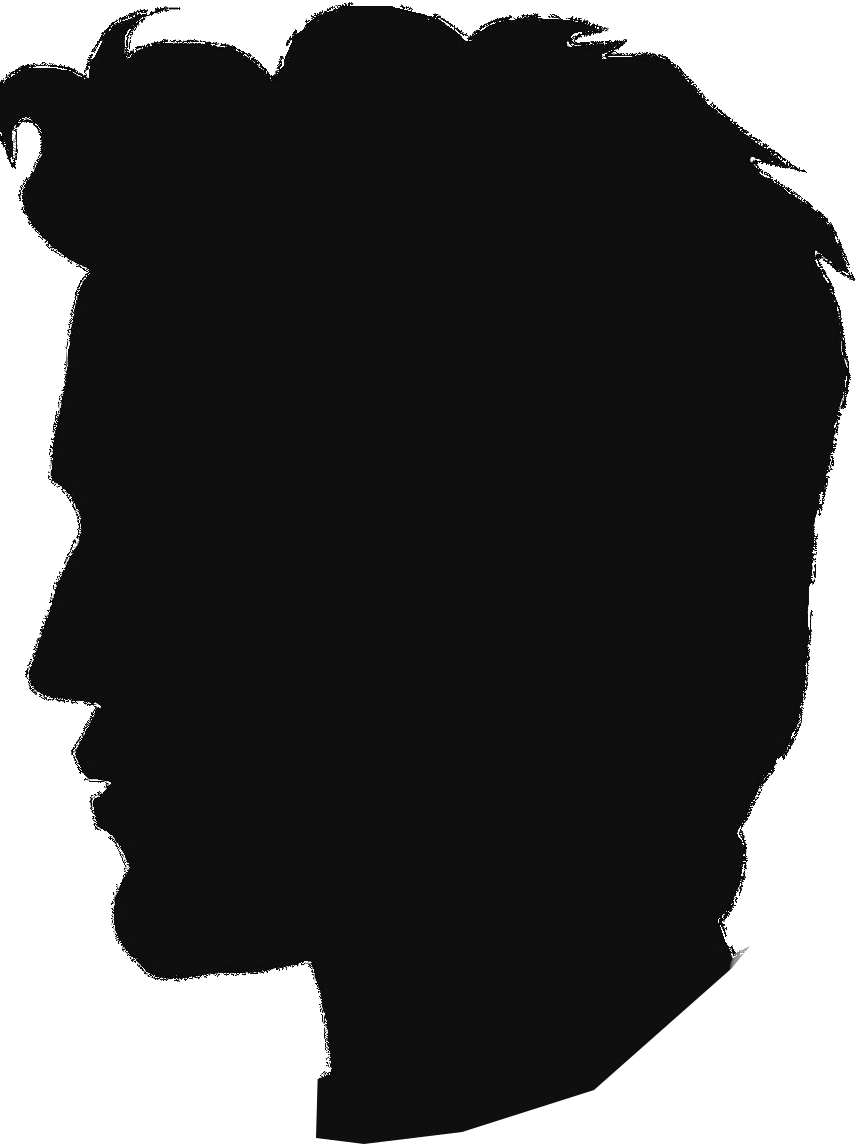}};
    \node[] (W) at (9,-2) {\textcolor{Wien}W};
 \end{tikzpicture}
 \caption{Simple Wigner's friend experiment: The source emits a quantum state $\ket{\phi}_S$, which is measured by the friend in the computational basis \{$\ket{0}_S, \ket{1}_S$\}. The result observed by the friend is stored in some memory register $\ket{\cdot}_F$ and she ascribes the respective product $\ket{i}_S\ket{\bm i}_F$, with $i \in \{ 0,1\}$, to herself and the system she measured. Wigner performs a measurement on both the system and the friend's memory register, which according to his unitary description is in state $\ket{\Phi}_{SF}$ that is, in general, a superposition of $\ket{0}_S\ket{\bm 0}_F$ and $\ket{1}_S\ket{\bm 1}_F$. The different state assignments to the joint system $S+F$ will lead to different probability assignments for Wigner's measurement.
 }
    \label{Simple Wigner}
\end{figure}
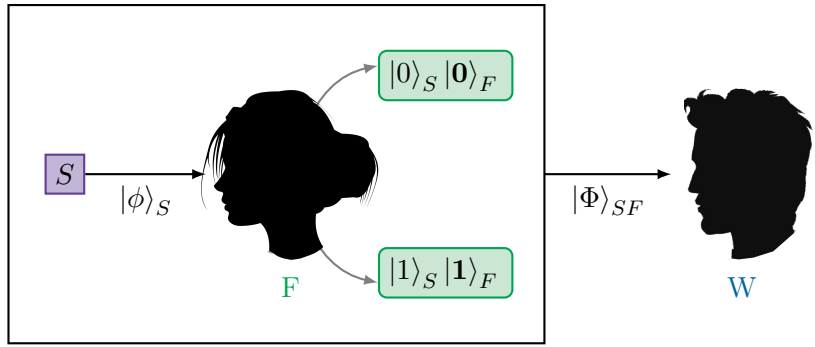

The original thought experiment features an observer --called Wigner's friend-- who measures a quantum system, as well as a so-called superobserver --Wigner-- who performs a joint measurement on the quantum system S and the friend F. Provided that the joint system S+F is sufficiently isolated, Wigner describes the friend's interaction with the system via unitary dynamics. To Wigner's friend, however, this interaction constitutes a measurement and she uses the collapse postulate after observing an outcome. These disagreeing descriptions of one and the same situation is called the Wigner's friend paradox, see Fig.~\ref{Simple Wigner}. Let the source emit a qubit in the state $\ket{\phi}_S=\alpha\ket{0}_S+\beta\ket{1}_S$, which is measured by the friend in the computational basis \{$\ket{0},\ket{1}$\}. The result she observes, $f\in \{ \bm0, \bm1\}$ is stored in her memory register $\ket{\cdot}_F$, which is supposed to correspond to the friend having a perception of the outcome $f$. Wigner then performs a measurement on the qubit and the friend's memory given by the states $\ket{1}_{SF}=a \ket{0,\bm0}_{SF}+b\ket{1,\bm1}_{SF}$ and $\ket{2}_{SF}=b^* \ket{0,\bm0}_{SF}-a^*\ket{1,\bm1}_{SF}$ and their orthogonal complement. Due to their different descriptions of the friend's measurement Wigner and his friend will assign different probabilities to the outcomes of Wigner's measurement. According to the friend after her measurement the qubit and her memory are either in state $\ket{0,\bm0}_{SF}$, which happens with probability $p(0)=|\alpha|^2$, or in state $\ket{1,\bm1}_{SF}$, which happens with probability $p(1)=|\beta|^2$. Hence, the friend  will assign the following probability to Wigner's results
\begin{equation}
{\color{Green}p^{F}(w)}= p^{clps}(w)= |\alpha|^2 |\braket{w|0,\bm0}|^2+|\beta|^2 |\braket{w|1,\bm1}|^2,
\label{p_clsp}
\end{equation}
where $\ket{w}$ is either $\ket{1}_{SF}$ or $\ket{2}_{SF}$. Wigner, however, assigns the state $\ket{\Phi}_{SF}=\alpha \ket{0,\bm0}_{SF}+\beta\ket{0,\bm0}_{SF} $ to the qubit and his friend's memory and, hence,  probabilities
\begin{equation}
{\color{Wien}p^{W}(w)}=  p^{uni}(w)= |\braket{w|\Phi}_{SF}|^2.
\label{p_clsp}
\end{equation}
More concretely, we obtain the following predictions for Wigner's measurement according to Wigner and his friend in the simple Wigner's friend experiment
\begin{equation}
\renewcommand{\arraystretch}{1.5}
  \begin{array}{l c|c c l c|c}
  \color{Wien} p^{W}(w):&1&2& \qquad \qquad \qquad &\color{Green} p^{F}(w):&1&2 \\ 
  \cline{2-3}\cline{6-7}
  &(\alpha a^*+\beta b^*)^2&(\beta a-\alpha b)^2&   & &|\alpha|^2|a|^2+ |\beta|^2 |b|^2& |\beta|^2 |a|^2+|\alpha|^2 |b|^2.
  \end{array} 
 \label{Pnorecords}
\end{equation}
Wigner originally argued in favor of the friend's description (and probability assignment) claiming that at the level of an observer, at the latest, unitary quantum theory must break down. Since then, however, the idea of observers in superposition has become accepted~\cite{deutsch1985quantum}, which begs the question of whether the disagreement between Wigner and his friend can be experimentally verified. As already discussed in~\cite{baumann2018formalisms,cavalcanti2021view,baumann2020wigner}, the unitary description of the friend's measurement is incompatible with the existence of a record revealing which result the friend observed before Wigner performs his measurement. Such a record destroys any coherences Wigner could reveal in his measurement. However, there are persistent records, which do not contain any outcome information of the friend's measurement and can, therefore, be present without destroying the coherence of an observer in superposition~\cite{deutsch1985quantum,baumann2020wigner}. In special cases the disagreement between Wigner and his friend becomes manifest in terms of contradicting records that both Wigner and his friend can access after the thought experiment has been performed. Note that, in these cases there is no persistent record of which result the friend observed at her measurement, since Wigner's measurement will, in general, alter the friend's perception of her measurement result, i.e. the result stored in the internal memory register~\cite{allard2020no}. \\%As shown in~\cite{allard2020no}, if one, nevertheless, considers a joint probability distribution of the friend's perception before and after Wigner's measurement, this probability distribution cannot agree with the unitary description endorsed by Wigner and be a convex linear function of the initial state. \\

\begin{figure}[hbt!]
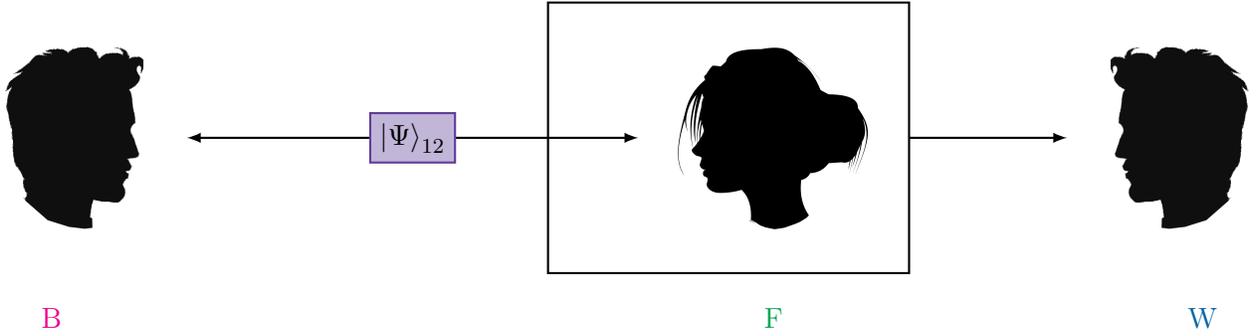

\begin{tikzpicture}[scale=1.2]

  \draw[<->,>=latex, thick,] (-2.5,0)--  (2.5,0);
  \node[fill=RoyalPurple!30, draw=RoyalPurple, thick] (s) at (0,0) {$\ket{\Psi}_{12}$};

  \draw[thick,] (5.5,1.5) rectangle (1.5,-1.5);
  \node[xscale=-1] (m) at (4,0) {\includegraphics[scale=0.0375]{friend2}};
    \node[] (F) at (4,-2) {\textcolor{Green}F};
  \draw[->,>=latex,thick] (5.5,0) -- (7.25 ,0);
  \node[] (M) at (8.5,0) {\includegraphics[scale=0.08]{Wigner}};
    \node[] (W) at (8.75,-2) {\textcolor{Wien}W};
  \node[xscale=-1] (M) at (-3.75,0) {\includegraphics[scale=0.08]{Wigner}};
    \node[] (B) at (-4,-2) {\textcolor{Magenta}B};
 \end{tikzpicture}
 \caption{Simplest extended Wigner's friend experiment: A bipartite state $\ket{\Psi}_{12}$ is emitted by the source. One subsystem is measured by the friend -- F -- in a simple Wigner's friend setup. Together with the subsystem she interacted with the friend is then measured by Wigner -- W. The other subsystem is measured by an additional observer Bob -- B. 
  }
    \label{Extended Wigner}
\end{figure}

The newest versions of the thought experiment combine multiple Wigner's friend scenarios with various non-locality proofs~\cite{haddara2022possibilistic,ormrod2022no,leegwater2022greenberger,Zukowski2020,Bong2020,brukner2018no,frauchiger2018quantum,brukner2017quantum}. Some of these extended Wigner's friend experiments rely on conflicting probability assignments of observers and superobservers similar to the simple Wigner's friend experiment above. This lead to closer investigations of how agents in these settings can make predictions and consistently reason about each other~\cite{vilasini2019multi,vilasini2022general}. Other extended Wigner's friend scenarios concern the joint probabilities of the results of observers and superobservers, similar to Bell's theorem. More concretely,  a set of assumptions called local friendliness --  namely, that the superobservers’ and observers’ results are both ``objective facts'', locality, freedom of choice, and universality of quantum theory, meaning observers can exist in superpositions of different observation states -- cannot all hold in extended Wigner's friend setups. In the simplest extended Wigner's friend scenario,  as depicted in Fig.~\ref{Extended Wigner}, a bipartite entangled state is shared between a Wigner's friend setup and one additional observer -- Bob. The violation of a CHSH-like-inequality between Wigner and Bob asserts that the local friendliness assumptions cannot hold simultaneously. \\

Consider the setup in Fig.~\ref{Extended Wigner}, where Wigner randomly chooses between the measurement that reveals which result his friend observed in her measurement $W_z = \proj{0,\bm0}_{2F}-\proj{1,\bm1}_{2F}$ and one projecting on the states $\ket{\Phi^{\pm}}_{2F}=1/\sqrt{2}(\ket{0,\bm0}_{2F}\pm \ket{1,\bm1}_{2F})$, i.e. $W_x=\ketbra{0,\bm0}{1,\bm1}_{2F} + \ketbra{1,\bm1}{0,\bm0}_{2F}$. Bob, on the other hand, performs measurements $B_z=1\sqrt{2}(\proj{0}+\ketbra{0}{1}+\ketbra{1}{0}- \proj{1})$ and $B_x=1\sqrt{2}(\proj{0}-\ketbra{0}{1}-\ketbra{1}{0}-\proj{1})$ on qubit 1. One  local friendliness inequality for this scenario is given by the following CHSH-expression 
\begin{equation}
\langle B_z \otimes W_z \rangle + \langle B_x \otimes W_z \rangle -\langle B_z \otimes W_x \rangle +\langle B_x \otimes W_x \rangle \leq 2.
\label{CHSH-LF}
\end{equation}
Other inequalities can be obtained by making an alternative choice for which expectation value is subtracted on the left hand side. If the source emits the singlet state $\ket{\psi^-}_{12}= 1/\sqrt{2}(\ket{0,0}_{12}-\ket{1,1}_{12})$ and the friend measures qubit 2 in the computational basis, we obtain the overall state 
\begin{equation}
\ket{\Psi} = \frac{1}{\sqrt{2}} \left(\ket{0}_1 \ket{0,\bm0}_{2F} - \ket{1}_1 \ket{1,\bm1}_{2F} \right)
\label{state-LF}
\end{equation}
on which Wigner and Bob perform their measurements. This give the following violation of the inequality in Eq.~\eqref{CHSH-LF}
\begin{equation}
\langle B_z \otimes W_z \rangle + \langle B_x \otimes W_z \rangle -\langle B_z \otimes W_x \rangle +\langle B_x \otimes W_x \rangle =4\cdot \frac{1}{\sqrt{2}}= 2\sqrt{2} > 2.
\label{CHSH-LFviolation}
\end{equation}
The violations of local friendliness inequalities have been confirmed experimentally in proof of principle experiments~\cite{Bong2020,proietti2019experimental}, where an additional qubit played the role of Wigner's friend. Such experiments arguably do not constitute genuine Wigner's friend experiments, since the interaction between two qubits does not satisfy many basic characteristics of an observation~\cite{brukner2021qubits}. In response to that, an extended Wigner's friend experiment involving a human level AI on a quantum computer playing the role of the friend has been proposed~\cite{wiseman2022thoughtful}. Such a friend would satisfy most qualitative features of an observer while operating fully unitarily by construction.\\

The rest of this article is structured as follows. In the main Section~\ref{Classical information and collapse} I consider communication between Wigner and his friend first by incorporating record systems that are not subject to Wigner's measurement in~\ref{Effective collapse}. Second, I introduce a (quasi) classical communication channel into the simple Wigner's friend experiment in~\ref{Partial collapse}. This allows to recover probabilities in agreement with collapse dynamics as well as with unitary dynamics and anything in between depending on the properties of the communication channel. In Section~\ref{Local friendliness inequalities and communication} I then consider the records as well as the communication channel in an extended Wigner's friend setup and discuss their implications on the violation of the local friendliness inequality presented above. Finally, the conclusions are summarize in Section~\ref{Conclusion}.

%%%%%%%%%%%%%%%%%%%%%%%%%%%%%%%%%%%%%%%%%%%%%%%%%%%%%%
 
\section{Classical information and collapse}
\label{Classical information and collapse}

As discussed in~\cite{deutsch1985quantum,zurek2018quantum,cavalcanti2021view} unitary dynamics is incompatible with simultaneously preserving a classical record of a measurement result. For Wigner's friend experiments this means that there cannot be a record revealing the friend's observed outcome. In general, records or messages that are not subject to Wigner's measurement influence the probabilities for Wigner’s results according to a unitary description of the setup.

\subsection{Effective collapse}
\label{Effective collapse}

If the friend produces a classical record revealing her observed result, also Wigner's description of the setup gives rise to the probabilities induced by collapse dynamics. Consider a record Hilbert space $\mathcal{H}_R$ with a fixed basis \{$\ket{r_i}$\}, which encodes the (quasi) classical messages Wigner receives from his friend. For example, in the simplest Wigner's friend experiment in Fig.~\ref{Simple Wigner} these messages could correspond to $\ket{r_0}=\ket{\text{``I saw $0$.''}}$ and $\ket{r_1}=\ket{\text{``I saw $1$.''}}$. A unitary description of the setup, then, leads to the following over all state
\begin{equation}
\ket{\Psi^r} = \alpha \ket{0,\bm0}_{SF}\ket{r_0}_R+\beta \ket{1,\bm1}_{SF}\ket{r_1}_R,
\label{state_classical}
\end{equation}
upon which Wigner performs his measurement. The probabilities for Wigner's measurement result are then given by
\begin{equation}
{\color{Wien}p^{W}(w)}= \Tr \left( \proj{w}_{SF} \proj{\Psi^r}\right),
\label{probW_classical}
\end{equation}
which is equal the ones assigned by the friend who uses collapse dynamics
\begin{equation}
\renewcommand{\arraystretch}{1.5}
  \begin{array}{l c|c}
  {\color{Wien} p^{W}(w)}={\color{Green} p^{F}(w)}:&1&2\\ 
  \cline{2-3} 
  &|\alpha|^2|a|^2+|\beta|^2|b|^2&|\beta|^2|a|^2+|\alpha|^2|b|^2.
  \end{array} 
 \label{Precords}
\end{equation}
Note that, tracing out the record system can be understood as Wigner ignoring the record of the friend's result. Yet the existence of such a message alone, even if Wigner does not know what it reads, effectively collapses the state of the system and the friend. When taking into account which result the friend observed we need to consider conditional probabilities. In the collapse description employed by the friend, this means either using the state $\ket{0,\bm0}$ or $\ket{1,\bm1}$ when calculating ${\color{Green} p^{F}(w]f)}= |\braket{w|f,\bm f}]^2$. In Wigner's unitary description we condition on the record by evaluating probabilities
\begin{align}
 {\color{Wien} p^W(w,j)}= \Tr{\proj{w}_{SF} \otimes \proj{r_j}_R \proj{\Psi^r}}=  p(j)\cdot {\color{Wien} p^W(w|j)}
\label{cond_prob(w)}
\end{align}
where $p(j)= \Tr(\mathds{1}\otimes \proj{r_j}_R \proj{\Psi^r})$ is the probability of message $r_j$ and we define $p(w|j)=0$ if  $p(j)=0$. This gives
\begin{equation}
\renewcommand{\arraystretch}{1.5}
  \begin{array}{l c|c c l c|c}
  {\color{Wien}p^W(w|0)}={\color{Green} p^{F}(w]f=0)}:&1&2& \qquad \qquad \qquad 
  &{\color{Wien}p^W(w|1)}={\color{Green} p^{F}(w]f=1)}:&1&2 \\ 
  \cline{2-3}\cline{6-7}
  & |a|^2& |b|^2&   & & |b|^2& |a|^2.
  \end{array} 
 \label{records_clps}
\end{equation}
if the record reveals which result the friend observed. Conversely, if the record space is only one dimensional, for example $\ket{r'}=\ket{\text{``I saw a definite outcome''}}$, the record factors out and 
\begin{equation}
\ket{\Psi^{r'}} = (\alpha \ket{0,\bm0}_{SF}+\beta \ket{1,\bm1}_{SF})\ket{r'}_R,
\label{state_DD}
\end{equation}
which preserves the disagreement between Wigner and his friend. Conditioning on the record is obsolete in this case and the probabilities ${\color{Wien}p^{W}(w)}= \Tr \left( \proj{w}_{SF} \proj{\Psi^{r'}}\right)\neq{\color{Green} p^{F}(w)}$ are again those in Eq.~\eqref{Pnorecords}. This is due to the fact that such a one dimensional record cannot reveal any outcome information of the friend's two-outcome measurement.\\

%In what follows we will consider the joint probabilities $p_m(f,w)$ of the entries in the friend's and Wigner's memory after Wigner's measurement, potentially conditioned on a certain message or record $m$. This can be understood as the information available to Wigner and his friend after they have performed the experiment. One can think of this as Wigner opening the box after the friend's measurement of the qubit followed by Wigner's own measurement(s) an now freely communicating with his friend. 

\subsection{Partial collapse}
\label{Partial collapse}

\begin{figure}[hbt!]
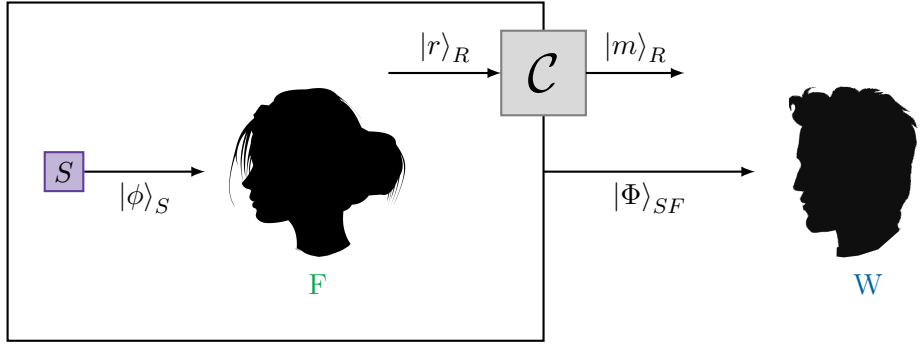

\begin{tikzpicture}[scale=0.75]

  \draw[thick,] (-4,-3) rectangle (5.5,3);
  \node[fill=RoyalPurple!30, draw=RoyalPurple, thick] (s) at (-3,0) {$S$};
  
  \node[xscale=-1] (m) at (1.5,0) {\includegraphics[scale=0.035]{friend2}};
  \node[] (F) at (1.5,-2) {\textcolor{Green}F};
 
   \draw[->,>=latex, thick,] (s) -- node [midway,below] {$\ket{\phi}_S$} (-0.5,0);
     
 \draw[thick,draw=gray, fill=gray!30] (4.75,2.5) rectangle (6.25,1);
   \node[] (C) at (5.5,1.75) {\LARGE$\mathcal{C}$};

 \draw[->,>=latex, thick] (2.75,1.75) --node [midway,above]{$\ket{r}_{R}$} (4.75,1.75);
 \draw[->,>=latex, thick] (6.25,1.75) --node [midway,above]{$\ket{m}_{R}$} (8,1.75);
 
  \draw[->,>=latex, thick] (5.5,0) --node [midway, below]{$\ket{\Phi}_{SF}$} (9.25,0);
  
  \node[] (M) at (11,0) {\includegraphics[scale=0.075]{Wigner}};
    \node[] (W) at (11.25,-2) {\textcolor{Wien}W};
    
 \end{tikzpicture}
 \caption{Communication channel $\mathcal{C}$ between Wigner and his friend: The friend can encode a message via a fixed basis \{$\ket{r}$\} of some record Hilbert space $\mathcal{H}_R$. The channel $\mathcal{C}$ takes these record states as input and produces a classical message, which can in principle be encoded in another basis \{ $\ket{m}$\} of the record Hilbert space. This freedom to choose different bases for the incoming and the outgoing messages allows for controlling how much outcome information the friend can send to Wigner via this channel.
 }
 \label{Simple Wigner CC}
\end{figure}

\begin{figure}[hbt!]
\centering
\includegraphics[width=0.95\linewidth]{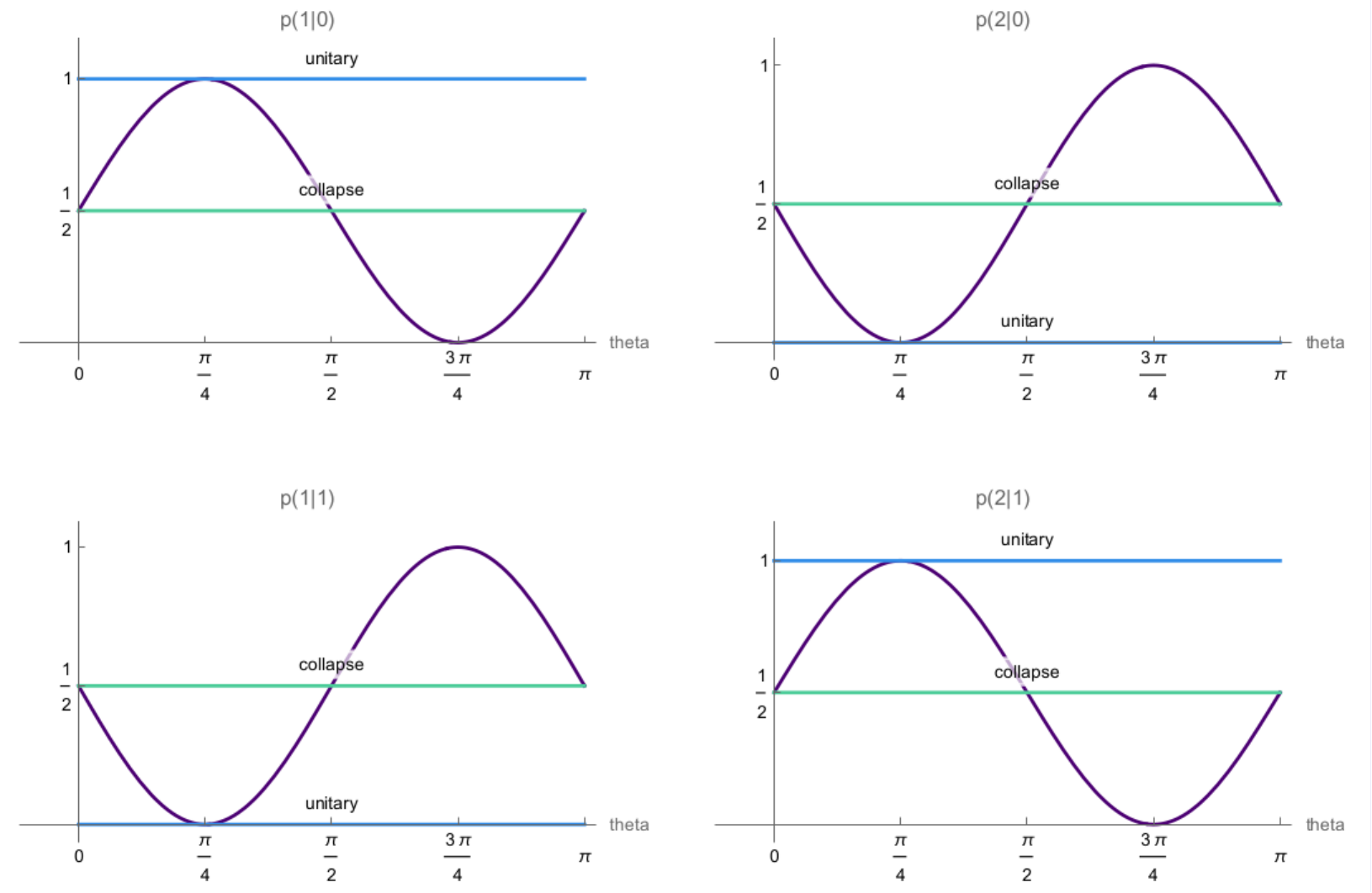} 
\caption{Partial collapse for communication between Wigner and his friend: If the source in the setup depicted in Fig.\ref{Simple Wigner CC} emits the state $\ket{\phi}= 1/\sqrt{2}(\ket{0}+\ket{1})$ and Wigner measures in the Bell-basis \{$\ket{\phi^+},\ket{\phi^-},\ket{\psi^+},\ket{\psi^-}$\} he will obtain results ``$+$'' and ``$-$'' corresponding to states $\ket{\phi^+}=1/\sqrt{2}(\ket{0,\bm 0}+\ket{1, \bm 1}$ and $\ket{\phi^-}1/\sqrt{2}(\ket{0,\bm 0}-\ket{1, \bm 1}$ respectively. For a communication channel $\mathcal{C}(\theta,\phi)$, and with $\phi=0$, the probabilities for Wigner's outcomes according to the unitary description of the setup depend on parameter $\theta$ as shown above. These probabilities (violet) vary between those corresponding to the effectively collapsed state (green) and the effectively unitary case (blue).
 }
 \label{Partial collapse_Simple Wigner}
\end{figure}

We now consider a more general scenario, where instead of directly exchanging messages, there is a (quasi) classical communication channel~\cite{holevo1998quantum} between Wigner and his friend, see Fig.~\ref{Simple Wigner CC}. Such a channel measures the incoming state and prepares a corresponding outcome in some fixed basis \{$\ket{n}$\} of the record Hilbert space $\mathcal{H}_R$
\begin{equation}
\mathcal{C}[\sigma ] := \sum_{m} \bra{m}\sigma\ket{m} \proj{m}.
\label{channel}
\end{equation}
Other than in Section~\ref{Effective collapse} the messages $\ket{m}$ sent to Wigner can now be encoded in a different basis of $\mathcal{H}_R$ than the records $\ket{r}$ the friend produces. This allows for these messages to only partially reveal which outcome the friend observed. The state Wigner performs his measurement on is now 
\begin{align}
\rho_{SFR} =& (\mathds{1} \otimes \mathcal{C} ) \proj{\Psi^r} = \sum_m |\alpha|^2|\braket{m|r_0}|^2 \rho^{00}_{SF} \otimes \proj{m}_R+ |\beta|^2|\braket{m|r_1}|^2 \rho^{11}_{SF} \otimes  \proj{m}\nonumber\\
&+ \alpha\beta^*\braket{m|r_0}\braket{r_1|m} \rho^{01}_{SF} \otimes  \proj{m}+ \alpha^*\beta \braket{m|r_1}\braket{r_0|m} \rho^{10}_{SF} \otimes  \proj{m}, \label{state_classChannel}
\end{align}
where 
\begin{align*}
\rho^{00}_{SF}&= |a|^2 \proj{1}_{SF}+|b|^2\proj{2}_{SF} + a^* b^* \ketbra{1}{2}_{SF}+ab\ketbra{2}{1}_{SF}, \\ 
\rho^{11}_{SF}&=|b|^2 \proj{1}_{SF}+|a|^2\proj{2} _{SF}- a^* b^* \ketbra{1}{2}_{SF} -ab\ketbra{2}{1}_{SF}, \\ 
\rho^{01}_{SF}&=a^*b( \proj{1}_{SF}-\proj{2}_{SF}) - a^* a^* \ketbra{1}{2}_{SF} + bb\ketbra{2}{1}_{SF}, \\
\rho^{10}_{SF}&=b^*a (\proj{1}_{SF}-\proj{2}_{SF}) + b^*b^* \ketbra{1}{2}_{SF}-aa\ketbra{2}{1}_{SF},
\end{align*}
with $\ket{1}_{SF}=a \ket{0,\bm0}_{SF}+b\ket{1,\bm1}_{SF}$, $\ket{2}_{SF}=b^* \ket{0,\bm0}_{SF}-a^*\ket{1,\bm1}_{SF}$ being the eigenstates corresponding to Wigner's measurement results ``1'' and ``2'' respectively. Analogous to the Eq.~\eqref{cond_prob(w)} in section~\ref{Effective collapse} we consider the conditional probabilities of Wigner's result $w$ given message $n$, i.e. $p(n) p(w|n)= \Tr(\proj{w}_{SF} \otimes \proj{n}_R \rho_{SFR})$, obtaining the following joint probabilities 
%put out by the channel, given by
%p(w,n)= \Tr{\proj{w}_{SF} \otimes \proj{n}_R \rho_{SFR}}= p(n)\cdot p(w|n),
%\label{cond_prob(w)_gen}
%\end{align}
%The joint probabilities are the following
\begin{equation}
\renewcommand{\arraystretch}{1.5}
  \begin{array}{l c|c}
&p(w=1,n)&p(w=2,n)\\ 
  \cline{2-3} 
  &|\alpha|^2|a|^2|\braket{n|r_0}|^2+|\beta|^2|b|^2|\braket{n|r_1}|^2
  &|\beta|^2|a|^2|\braket{n|r_0}|^2+|\alpha|^2|b|^2|\braket{n|r_1}|^2\\
  &+\alpha \beta^*a^*b \braket{n|r_0}\braket{r_1|n}+\alpha^* \beta a b^* \braket{n|r_1}\braket{r_0|n}
  &-\alpha \beta^*a^*b \braket{n|r_0}\braket{r_1|n}-\alpha^* \beta a b^* \braket{n|r_1}\braket{r_0|n}.
  \end{array} 
 \label{Precords_gen}
\end{equation}
The overlaps $\braket{n|r_i}$ indicate how much outcome information Wigner can obtain from the channel output message $n$. If the basis \{$\ket{n}$\} is the same as \{$\ket{r_i}$\}, i.e.$\braket{n|r_i}=\delta_{ni}$, the message perfectly reveals which result the friend observed and $p(n=0)=|\alpha|^2$, $p(n=1)=|\beta|^2$. In this case we recover the collapse behavior in Eq.\eqref{records_clps}, as discussed in Section \ref{Effective collapse}. However, if the two bases are mutually unbiased, for example $\braket{0|r_i}=\braket{1|r_0}=1/\sqrt{2}=-\braket{1|r_1}$, the records reveal no outcome information about the friend's measurement and $p(n)=1/2$ for both $n$. In this case, we obtain probabilities in accordance with a unitary description for each of the records
\begin{equation}
\renewcommand{\arraystretch}{1.5}
  \begin{array}{l c|c c l c|c}
  p(w|0):&1&2& \qquad \qquad \qquad &p(w|1):&1&2 \\ 
  \cline{2-3}\cline{6-7}
 &(\alpha a^*+\beta b^*)^2&(\beta a-\alpha b)^2&   &  &(\alpha a^*-\beta b^*)^2&(\beta a+\alpha b)^2. 
  \end{array} 
 \label{records_uni}
\end{equation}
Note that the phase shift between the expressions for the two messages does not allow for simply adding them when wanting to preserve the resemblance to the unitary description without records. Evaluating $p(w|0)+p(w|1)$ corresponds to tracing out the record system and, as already mentioned in Section~\ref{Effective collapse}, gives the collapse probabilities in Eq.~\eqref{Precords}. In general, we can express the messages Wigner receives in the basis of the records generated by the friend as follows
\begin{align}
\ket{0}&= \cos{\theta} \ket{r_0}+ e^{i\phi}\sin{\theta} \ket{r_1}
\label{message_state0} \\
\ket{1}&= e^{-i\phi}\sin{\theta} \ket{r_0}- \cos{\theta}\ket{r_1},
\label{message_state1}
\end{align}
where one can think of $\theta, \phi$ as variable parameters of the communication channel $\mathcal{C}$. Hence, the joint probabilities for Wigner's measurement result and message $n$ are given by
\begin{equation}
\renewcommand{\arraystretch}{1.5}
  \begin{array}{l c|c}
p(w,0):&1&2\\ 
  \cline{2-3} 
  &|\alpha|^2|a|^2\cos^2(\theta)+|\beta|^2|b|^2\sin^2(\theta)
  &|\beta|^2|a|^2 \cos^2(\theta)+|\alpha|^2|b|^2\sin^2(\theta)\\
  &+ \sin(2\theta)(\alpha \beta^*a^*b e^{i\phi}+\alpha^* \beta a b^* e^{-i\phi})  
  & - \sin(2\theta)(\alpha \beta^*a^*b e^{i\phi}+\alpha^* \beta a b^* e^{-i\phi})  
  \end{array} 
 \label{Precords_thetaphi0}
\end{equation}
and
\begin{equation}
\renewcommand{\arraystretch}{1.5}
  \begin{array}{l c|c}
p(w,1):&1&2\\ 
  \cline{2-3} 
  &|\alpha|^2|a|^2\sin^2(\theta)+|\beta|^2|b|^2\cos^2(\theta)
  &|\beta|^2|a|^2\sin^2(\theta)+|\alpha|^2|b|^2\cos^2(\theta)\\
  &- \sin(2\theta)(\alpha \beta^*a^*b e^{i\phi}+\alpha^* \beta a b^* e^{-i\phi})
  & + \sin(2\theta)(\alpha \beta^*a^*b e^{i\phi}+\alpha^* \beta a b^* e^{-i\phi}).
  \end{array} 
 \label{Precords_thetaphi1}
\end{equation}
Varying $\theta$ and $\phi$ then allows for smoothly changing between the effectively collapsed case in Eq.~\eqref{records_clps} and the effectively unitary case in Eq.~\eqref{records_uni}, see Fig.~\ref{Partial collapse_Simple Wigner} for a simple example. Note that, for the partially collapsed scenarios there is still a   paradoxical situation, since the friend when describing her measurement via collapse dynamics will always assign the probabilities in Eq.~\eqref{records_clps}. Wigner's probability assignments, namely those derived from Eqs.~\eqref{Precords_thetaphi0}-\eqref{Precords_thetaphi1}, will always differ from his friend's unless the messages fully reveal which outcome the friend observed. 

%%%%%%%%%%%%%%%%%%%%%%%%%%%%%%%%%%%%%%%%%%%%%%%%%%%%%%

\subsection{Local friendliness inequalities and communication}
\label{Local friendliness inequalities and communication}

Analogous to the simple Wigner's friend experiment the presence of classical records of the friend's observed result effectively collapses the state Wigner performs his measurement on also in the extended Wigner's friend setup depicted in Fig.~\ref{Extended Wigner}. This, in turn, prevents the violation of the local friendliness inequality presented in Section~\ref{Introduction}. More concretely, using state
\begin{equation}
\ket{\Psi^r} = \frac{1}{\sqrt{2}} \left(\ket{0}_1 \ket{0,\bm0}_{2F}\ket{r_0}_R - \ket{1}_1 \ket{1,\bm1}_{2F}\ket{r_1}_R \right)
\label{state-LF_class}
\end{equation}
instead of the one in Eq.~\eqref{state-LF} leads to the following expression for the CHSH-like local friendliness inequality
\begin{equation}
\langle B_z \otimes W_z \rangle + \langle B_x \otimes W_z \rangle -\langle B_z \otimes W_x \rangle +\langle B_x \otimes W_x \rangle = \frac{1}{\sqrt{2}}+\frac{1}{\sqrt{2}}+0+0=\sqrt{2}<2,
\label{CHSH-LFclass}
\end{equation}
which means that none of the local friendliness assumptions needs to be rejected. This is due to the fact that the presence of the records, revealing which result the friend observed, effectively collapse the state in Eq.~\eqref{state-LF_class} to
\begin{equation}
\Tr_R(\proj{\Psi^r}) = \frac{1}{2} \left(\proj{0}_1 \proj{0,\bm0}_{2F} + \proj{1}_1 \proj{1,\bm1}_{2F} \right),
\label{eff_state-LF}
\end{equation}
which means that any expectation value containing Wigner's $W_x$-measurement vanishes. Note that, this is also true when we condition on the friend's observed result, meaning that we use either $\proj{0}_1 \proj{0,\bm0}_{2F}$ or $\proj{1}_1 \proj{1,\bm1}_{2F}$ as the effective state. \\

\begin{figure}[hbt!]
\centering
\includegraphics[width=0.65\linewidth]{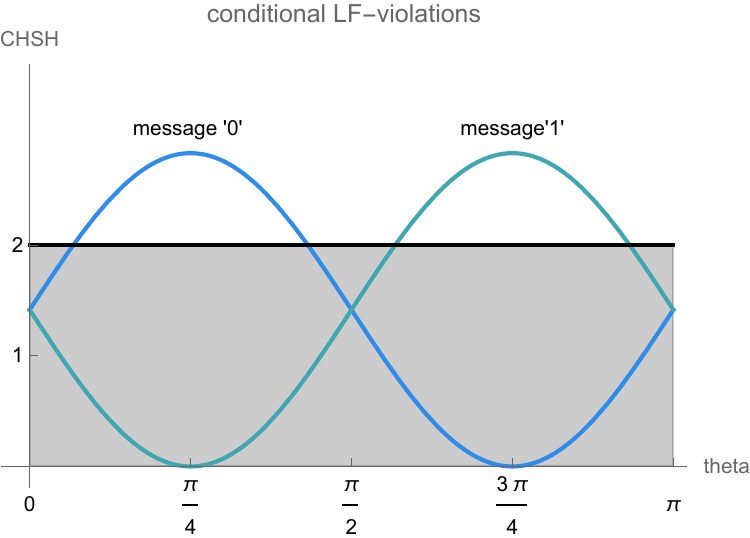} 
\caption{CHSH-values for the extended Wigner's friend setup with communication: The conditional expectation values $\langle B\otimes W\rangle^{|n}$ for the extended Wigner's friend experiment in Fig.~\ref{Extended Wigner} give CHSH-values depend on the communication channel parameters $\phi$ and $\theta$. For $\phi=0$ and $\theta \in [0,\pi]$, the expression given by conditioning on message ``$0$'' are shown in blue, the one corresponding to message ``$1$'' is depicted in green. There are values of $\theta$ where neither of the conditional CHSH-values lies above the local friendliness threshold of $2$, meaning no violation occurs. Whenever the CHSH-like local friendliness inequality is violated for one of the messages, when conditioning on the other message the local friendliness inequality is satisfied.
 }
 \label{LFvalue-extended_Wigner}
\end{figure}

We now, again, consider a general communication channel $\mathcal{C}$ between the friend and Wigner, as depicted in Fig.~\ref{Simple Wigner CC}, also for this extended setup. Starting from the state in Eq.~\eqref{state-LF_class} we now let the channel act on the register space and obtain the state
\begin{align}
\label{state-LF_CCgen}
\rho'_{12FR}&=(\mathds{1}\otimes \mathcal{C}) \left(\proj{\Psi^r}\right) \\
&= \frac{1}{2} \Big(\proj{0}_1 \otimes \proj{0,\bm0}_{2F}\otimes \mathcal{C}(\proj{r_0}) 
- \ketbra{0}{1}_1\otimes \ketbra{0,\bm 0}{1,\bm 1}_{2F}\mathcal{C}(\ketbra{r_0}{r_1}) \nonumber \\
& \quad- \ketbra{1}{0}_1\otimes \ketbra{1,\bm 1}{0,\bm 0}_{2F}\mathcal{C}(\ketbra{r_1}{r_0}) 
+\proj{1}_1 \otimes \proj{1,\bm1}_{2F}\otimes \mathcal{C}(\proj{r_1})\Big), \nonumber
\end{align}
upon which Wigner and Bob perform their measurements. The action of the classical channel on the records, again, gives terms of the form
\begin{align}
\sum_m \braket{m|r_i}\braket{r_j|m} \proj{m}, 
\end{align}
which, if there is no conditioning on the message lead to the effective collapse discussed above regardless of the properties of the channel, i.e. parameters $\phi, \theta$, since 
\begin{align}
\Tr \left(\sum_m \braket{m|r_i}\braket{r_j|m} \proj{m}\right) = \sum_m \braket{m|r_i}\braket{r_j|m} =\braket{r_j|r_i} =\delta_{ij}. 
\end{align}
Similar to the probabilities for Wigner's outcome in the simple Wigner's friend setup, we can now define the expectation values for the measurements of Bob and Wigner, conditioned on the message $n$ put out by the classical channel $\mathcal{C}$ as follows
\begin{equation}
\langle B \otimes W\rangle^{|n} := 
\begin{cases} 
          \frac{1}{p(n)} \Tr \left( B \otimes W\otimes \proj{n} \cdot  \rho \right) & \text{for } p(n)> 0 \\
          \qquad \qquad\qquad \qquad 0 & \text{for } p(n)=0,
       \end{cases}
\label{Def_condexpv}
\end{equation}
where the probability for the messages is now given by $p(n)= \Tr \left( \mathds{1}\otimes \proj{n} \cdot  \rho'_{12FR} \right)= 1/2(\cos^2(\theta)+\sin^2(\theta))=1/2$. Plugging the state in Eq.~\eqref{state-LF_CCgen} into this expression, then gives
\begin{align}
\langle B \otimes W \rangle^{|n} =& \Big( \braket{0|B|0} \braket{0,\bm 0 |W| 0, \bm 0} |\braket{n|r_0}|^2 + \braket{1|B|1} \braket{1,\bm 1 |W| 1, \bm 1} |\braket{n|r_1}|^2 \label{Expectaionvalue_m} \\
 & \quad- \braket{1|B|0} \braket{1,\bm 1 |W| 0, \bm 0} \braket{n|r_0}\braket{r_1|n}
 - \braket{0|B|1} \braket{0,\bm 0 |W| 1, \bm 1} \braket{n|r_1}\braket{r_0|n}
\Big). \nonumber
\end{align}
For the settings of Bob and Wigner presented in Section~\ref{Introduction} we obtain the conditional expectation values 
\begin{align*}
\langle B_z \otimes W_z \rangle^{|n} &=\frac{1}{\sqrt{2}}( |\braket{n|r_0}|^2+|\braket{n|r_1}|^2)=\frac{1}{\sqrt{2}}=\langle B_x \otimes W_z \rangle^{|n} ,\\
\langle B_z \otimes W_x \rangle^{|n} =& -\frac{1}{\sqrt{2}}(\braket{n|r_0}\braket{r_1|n}+\braket{n|r_1}\braket{r_0|n})
=-\langle B_x \otimes W_x \rangle^{|n}.
\end{align*}
Hence, when conditioned on the message $n$ the local friendliness inequality in Eq.~\eqref{CHSH-LFclass} becomes
\begin{align}
\label{CHSH-LF_CC0}
\langle B_z \otimes W_z \rangle^{|0} + \langle B_x \otimes W_z \rangle^{|0} -\langle B_z \otimes W_x \rangle ^{|0}+\langle B_x \otimes W_x \rangle^{|0} &= \sqrt{2}+\sqrt{2}\cdot \cos(\phi)\sin(2\theta), \\
\label{CHSH-LF_CC1}
\langle B_z \otimes W_z \rangle^{|1} + \langle B_x \otimes W_z \rangle^{|1} -\langle B_z \otimes W_x \rangle ^{|1}+\langle B_x \otimes W_x \rangle^{|1} &= \sqrt{2}-\sqrt{2}\cdot \cos(\phi)\sin(2\theta).
\end{align}
where the term $\cos(\phi)\sin(2\theta)$ is determined by the properties of the communication channel. If the messages perfectly reveal which result the friend observed, i.e. $\phi=k\cdot\pi$ and $\theta=l\cdot \pi/2$, the channel dependent term, which corresponds to the expectation values $\langle B \otimes W_x \rangle^{|n}$, vanishes and we obtain the expression in Eq.~\eqref{CHSH-LFclass} for both messages. If the messages reveal no outcome information about the friend's measurement, i.e. $\theta=l\cdot \pi/4$ and $\phi=k\cdot\pi/2$, conditioning on one of the two messages gives the maximal violation of $2\sqrt{2}$. Since the channel dependent term smoothly varies in the interval $[-1,1]$, one can obtain all values from $0$ to $2\sqrt{2}$ for the CHSH-like local friendliness inequality by controlling the channel parameters $\phi$ and $\theta$.  Note that, due to the different signs for the two messages, whenever the conditional expectation values for one message violate local friendliness, the conditional expectation values for the other message do not violate the inequality, compare Fig.~\ref{LFvalue-extended_Wigner}. This is similar to the conditional probabilities $p(w|n)$ for the simple Wigner's friend experiment discussed in Section~\ref{Partial collapse}. There due to the different signs in $p(w|0)$ and $p(w|1)$ these probabilites can exactly reproduce those according to unitary dynamics only for one of the two messages.  

%%%%%%%%%%%%%%%%%%%%%%%%%%%%%%%%%%%%%%%%%%%%%%%%%%%%%%

\section{Conclusions}
\label{Conclusion}

We considered the possibility of communication between Wigner and his friend in both the simple Wigner's friend scenario and the simplest extended Wigner's friend setup that allows for the violation of local friendliness inequalities. As we showed explicitly a classical message revealing which result the friend observed during her measurement effectively collapses the state of her and the system she measured also in the unitary description employed by Wigner. For the simple Wigner's friend experiment, this means that the probabilities for Wigner's measurement result according to the unitary description of the setup are the same as those corresponding to collapse dynamics. For the extended Wigner's friend setup such records and the corresponding effective collapse prevent the violation of local friendliness inequalities. \\
We further considered the more general scenario of a (quasi) classical communication channel between Wigner and his friend. In that case, the records the friend produces are the input to the channel while the messages Wigner receives are the output of that channel. Provided that the friend's records encode which outcome she observed, how much of that which-outcome information is revealed by the output now depends on the properties of the channel. Controling the channel parameters then allows for gradually changing between collapse and unitary behavior for Wigner's friend experiments. In case of the simple Wigner's friend experiment this means that the probabilities based on the unitary description of the setup, associated with Wigner, will gradually approach those assigned by the friend based on the collapse description. For the extended Wigner's friend scenario, the channel properties determine whether and to what extent local friendliness inequalities can be violated.\\
Both the recovery of probabilities corresponding to unitary dynamics without records and the maximum violation of local friendliness inequalities not only occur when the messages Wigner receives  contain no information about which outcome the friend observed, but also require conditioning on the messages. Simply ignoring the messages put out by the channel always leads to a full effective collapse. This can be understood as showing that Wigner's unitary description does not just signify his ignorance about which result his friend observed.

\newpage

\bibliography{bib}{}
\bibliographystyle{ieeetr}
\clearpage

%%%%%%%%%%%%%%%%%%%%%%%%%%%%%%%%%%%%%%%%%%%%%%%%%%%%%
\begin{appendix}

\end{appendix}

\end{document}